\documentclass[12pt]{article}
\makeatletter
\@addtoreset{equation}{section}
\makeatother

\usepackage[pdftex]{graphicx}
\usepackage{subfigure}
\usepackage{array}
\usepackage{amsmath}

\usepackage{rotating}
\usepackage{longtable}
\usepackage{array}
\usepackage{geometry}
\usepackage{color}
\usepackage{pdfpages}
\usepackage{multicol}
\usepackage{amssymb}
\usepackage{amsthm}
\usepackage{bm}
\usepackage{mathtools}
\usepackage{fancybox}
\usepackage{algorithm}
\usepackage{multirow}
\usepackage[affil-it]{authblk}
\usepackage{setspace}
\usepackage{pdflscape}
\usepackage{afterpage}
\usepackage{capt-of}
\usepackage{lipsum}
\allowdisplaybreaks

\DeclarePairedDelimiter\abs{\lvert}{\rvert}%

\usepackage{xcolor}
\renewcommand{\raggedright}{\leftskip=0pt \rightskip=0pt plus 0cm}
\title{Functional time series prediction\\ under partial observation of the future curve}

\author[1]{Shuhao Jiao\thanks{shuhao.jiao@kaust.edu.sa}}
\author[2]{Alexander Aue\thanks{aaue@ucdavis.edu}}
\author[1]{Hernando Ombao\thanks{hernando.ombao@kaust.edu.sa}}

\affil[1]{Statistics Program,  KAUST, Saudi Arabia}
\affil[2]{Department of Statistics, UC Davis, USA}

\date{}

\begin{document}
	\maketitle			
	\setlength\parindent{0pt}
	\setlength{\parskip}{1em}
	\theoremstyle{definition}
	\newtheorem{theorem}{Theorem}
	\newtheorem{lemma}{Lemma}
	\newtheorem{remark}{Remark}
	\newtheorem{Definition}{Definition}
	
\begin{abstract}
This paper tackles one of the most fundamental goals in functional time series analysis which is to provide reliable predictions for future functions. Existing methods for predicting a complete future functional observation use only completely observed trajectories.  We develop a new method, called partial functional prediction (PFP), which uses both completely observed trajectories and partial information (available partial data) on the trajectory to be predicted. The PFP method includes an automatic selection criterion for tuning parameters based on minimizing the prediction error, and the convergence rate of the PFP prediction is established. Simulation studies demonstrate that incorporating partially observed trajectory in the prediction outperforms existing methods with respect to mean squared prediction error. The PFP method is illustrated to be superior in the analysis of environmental data and traffic flow data. \\
	

\end{abstract}
	
\section{Introduction}

Functional data is collected in sociological, environmental, transportation, biological and clinical research. In this paper, we analyze the daily trajectories of the pollutant PM10  in Graz, Austria displayed in Figure 1. These are fine particulate matter with diameter less than 10 micrometers, measured in $\mu$g$/m^{3}$. The task in this paper is to develop a new method that predicts the unobserved part of future trajectories using all the past daily trajectories and the partially observed trajectory of the curve to be predicted. 
\begin{figure}[!h]
\centering
\includegraphics[scale=0.5]{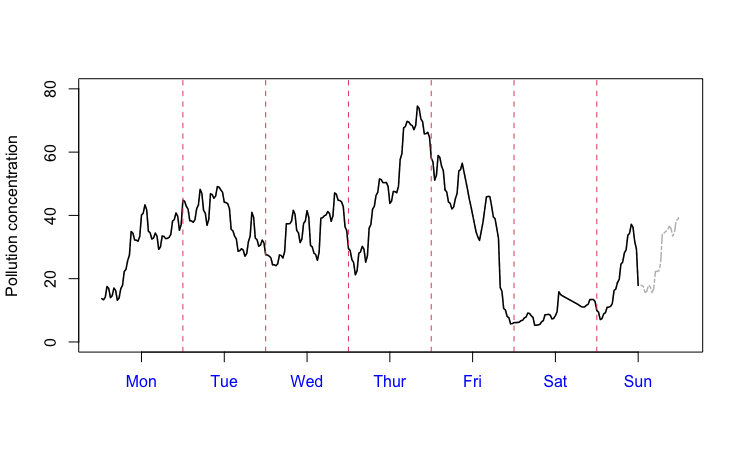}
\caption{Daily trajectories of the PM10 concentration ($\mu $g$m^{-3}$) over one week. The dotted grey line represents the unobserved part of Sunday's trajectory.}
\end{figure}
As noted, functional data are often collected over many natural consecutive time intervals. The PM10 dataset discussed above consists of many daily trajectories. One of the interesting aspects of functional time series is that the trajectories may share similar behavior and that these trajectories may be temporally correlated. Functional time series data is a time series of functions which may be described by $(Y_k(t) \ \colon \ k\in\mathbb{N})$, $\mathbb{N}$ denoting the integers, where the observations in the sequence are realizations of random functions $Y_k(t)$, and $t$ is in some domain 
$\mathcal{U}$ which here taken to be the unit interval $[0, 1]$. The object $(Y_k(t) \ \colon \ k \in\mathbb{N})$ will be referred to as a functional time series. 
	
Prediction methods using only completely observed curves have been discussed in the literature. However, one limitation of these methods is that they are not tailored for the case where there is partially observed data that is available for predicting the remaining part of the curve. Such existing methods often focus on the functional auto-regressive model (FAR).  In Bosq (2000),  a one-step ahead predictors based on a functional form of the Yule--Walker equations for FAR($1$)  processes is derived. Non-parametric kernel predictors were proposed in Besse, Cardot and  Stephenson (2000). FAR($1$) curve prediction based on linear wavelet methods were developed in Antoniadis, Paparoditis and Sapatinas (2006). Another line of interesting research in Kargin and Onatski (2008) is the predictive factor method which seeks to replace functional principal components with directions that are most relevant for predictions. Didericksen, Kokoszka and Zhang (2012) evaluated several competing prediction models in a comparative simulation study and report that the Bosq (2000) 
method has the best overall prediction performance based on mean squared error (MSE) and averaged distance. Aue, Dubart Norinho and H\"ormann (2015) proposed a prediction method that first projects the random functions onto the space spanned by the first few functional principal components. A VAR model is then used  to predict the truncated eigenscores of the future curve and then transform the predicted eigenscores to functional data by truncated Karhunen-Lo\'eve transformation. In Baranowski and Fryzlewicz (2020), the authors proposed the AMAR method for univariate time series where the AR coefficients are assumed to be constant over different segments. Existing full-curve prediction methods for functional time series only incorporate the dynamics across functions, but not the intra-function information. Thus partial observation is not utilized to improve the prediction. To overcome this limitation, we will propose a method that will incorporate information both across and within trajectories.

In contrast to complete curve prediction, our proposed partial functional prediction (PFP) method yields predictions that 
utilize both completely observed and partially observed trajectory. Fully functional regression methods for updating time series prediction have been proposed in Chiou (2012), and Yao and Shen (2017).  Motivated by the need to predict traffic flow, functional mixture methods were developed which combine fully functional regression, functional clustering, and discrimination. Shang (2017) also considered the fully functional regression method and implemented a moving block approach to update functional time series prediction. However, these approaches have some limitations. The moving block method is developed for prediction over the full interval $\mathcal{U}$, but not for the subinterval of the unobserved partial trajectory. Fully functional regression methods do not take the cross-curve information into account, which would be problematic when there existed strong cross-curve correlation. The PFP prediction method uses all available data -- both complete trajectories and partial trajectories. In contrast to methods that use only complete curves, the PFP method has the added flexibility because it can update the prediction according to different time of the interval $\mathcal{U}$. 
Consequently,  by incorporating partially observed trajectory, the prediction error over the forecasting time interval is 
anticipated to be smaller. 
	
The algorithm for the proposed prediction method is a stepwise procedure and can be summarized as follows. 
For smooth trajectories, the curves are decomposesd into two parts:
\[
Y_k(t)=S_k(t)+\epsilon_k(t),\hspace{0.7cm}  k=1,2,\ldots,
\]
where $S_k(t)$ is the random signal function, and $\epsilon_k(t)$ is the zero-mean $i.i.d$ innovation function across $k$. 
An example of this decomposition is FAR($p$) process, where $S_k(t)=\sum\limits_{h=1}^p\Phi_h(Y_{k-h})$, $\Phi_h$'s are the functional autoregressive operators and $Y_k$'s are zero-mean random functions without loss of generality. The innovation functions $(\epsilon_k(t)\colon k\in\mathbb{N})$ are assumed to be $i.i.d$ for the identifiability of signals and innovations. For $\tau\in (0,1]$, assume that the sub-function over the interval $[0,\tau]$, denoted by $Y_{n+1}|_{[0,\tau]}$, has already been observed and that the goal is to predict $Y_{n+1}|_{(\tau,1]}$. The first step in the proposed method is to use existing functional time series models (e.g.~FAR models) to calculate the fitted functions $\widehat{Y}_k(t)$ for all $t\in [0,1]$, and obtain the residual functions $\tilde{\epsilon}_k(t)=Y_k(t)-\widehat{Y}_k(t)$.  The  residual functions are then separated into two segments $\tilde{\epsilon}_k|_{[0,\tau]}$ and $\tilde{\epsilon}_k|_{(\tau,1]}$ at the current time $\tau$ and then the fully functional regression of $\tilde{\epsilon}_k|_{(\tau,1]}$ on $\tilde{\epsilon}_k|_{[0,\tau]}$ is employed to predict $\epsilon_{n+1}|_{(\tau,1]}$. The predicted function ${\hat{\epsilon}}_{n+1}|_{(\tau,1]}$ is then used to update the prediction of the $(\tau,1]$ block of the $(n+1)$-th function. The final prediction is derived to be $\widehat{Y}^u_{n+1}|_{(\tau,1]}=\widehat{Y}_{n+1}|_{(\tau,1]}+{\hat{\epsilon}}_{n+1}|_{(\tau,1]},$ which is the summation of predictions at each step, where $\widehat{Y}_{n+1}$ is the full-curve prediction. In the case where the observed trajectories are noisy, they are decomposed into three parts:
\[
Y_k(t)=S_k(t)+\epsilon_k(t)+e_k(t),\hspace{0.7cm}  k=1,\ldots,n.
\]
In addition to the two steps, a third step is included to extract the dependence information in the random error $e_k(t)$ 
which represents short-term dynamics (i.e., correlation across $t$ within a trajectory). Note that by properly accounting for the dependence between and across trajectories, the proposed PFP method achieves a non-trivial reduction on prediction error.
	
The remainder of this paper is organized as follows. Section 2 reviews the functional autoregressive model which is widely applied for ``next-interval'' prediction, and also the fully functional linear regression model and its application in partial functional prediction. In this section, limitations of the these methods are also discussed. In Section 3, we develop the prediction algorithm, and derive the convergence rate of the partial prediction. Simulation studies are discussed in Section 4.  In Section 5, the PFP method is validated in the analysis of PM10 concentration and also traffic flow trajectories. Section 6 concludes the article. Additional simulation results and technical details are included in the supplementary material.
	
\section{Functional Models for Prediction}
The two popular classes of models for analyzing correlated functional data are functional autoregressive model (FAR) and fully functional regression model. FAR models have been employed for analyzing dependent functional time series data. Fully functional regression models have been used to uncover linear associations between two sets of functions.  In this paper,  we develop the  partial functional prediction (PFP) method which is inspired by the principles of these two approaches. 

\subsection{Preliminaries}
Let $(Y_k(t) \colon k \in \mathbb{N})$ be an arbitrary stationary functional time series satisfying the following assumptions:
\begin{itemize}
\item[(A.1)] All random functions are defined on a common probability space. We assume the observations of $(Y_k(t) \colon k \in \mathbb{N})$ are elements of the Hilbert space $H = L^2[0, 1]$ equipped with the inner product $\langle x, y\rangle = \int_0^1 x(t)y(t) dt$, and $E\{\int_0^1 Y_k^2(t)dt\} < \infty$. 
\item[(A.2)] The mean function $\mu(t)$ and the covariance operator $C(x)$ of $(Y_k(t) \colon k \in \mathbb{N})$ are well defined, and are defined as $\mu(t)=E\{Y(t)\}\mbox{ and }C(x)(t)=\int_{0}^{1}c(t,s)x(s)ds$, where $c(t,s)=E\{Y(t)Y(s)\}$. By the Mercer's theorem, assume $C(x)$ admits the decomposition $C(x)=\sum\limits_{j=1}^{\infty}\lambda_j\langle v_j,x\rangle v_j,$ where $(\lambda_j\colon j\in\mathbb{N})$ are the eigenvalues in strictly descending order and $(v_j \colon j \in\mathbb{N})$ are the corresponding normalized eigenfunctions, so that $C(v_j) = \lambda_jv_j$ and $\|v_j\| = 1$. 
\end{itemize}

The set $(v_j(t) \colon j \in\mathbb{N})$ forms orthonormal bases of $L^2[0, 1]$. Then, by the Karhunen-Lo\`eve theorem, $Y_k(t)$ allows for the  representation $Y_k(t) = \mu(t)+\sum\limits_{j=1}^{\infty}\langle Y_k-\mu, v_j\rangle v_j(t).$ The coefficients $\langle Y_k-\mu, v_j\rangle$ in this expansion are called the fPC scores of $Y_k(t)$. Consider now the setting where $Y_1(t),\ldots,Y_n(t)$ are fully-observed functional time series with unknown mean function $\mu(t)$ and covariance operator $C(x)(t)$. Here, the estimators of $\mu(t)$ and $C(x)(t)$ are 
$$\hat{\mu}_n(t)=\frac{1}{n}\sum_{k=1}^nY_k(t), \qquad \widehat{C}_n(x)=\frac{1}{n}\sum_{k=1}^n\langle Y_k - \hat{\mu}_n, x\rangle(Y_k - \hat{\mu}_n), \qquad x\in H.$$
	
\begin{remark}
In (A.1), all functions are defined in the same probability space. Thus, there is common information from all the functions in the random sample that has to be extracted. The eigenfunctions $(v_j(t)\colon j\in\mathbb{N})$ in Assumption (A.2) form a series of orthonormal bases for the space $H$. Compared with other orthonormal basis, the eigenfunctions give the best approximation 
of functions for a given fixed number of basis functions. Under the PFP method, the prediction of functional time series is derived in the subspace spanned by the first few eigenfunctions associated with the largest eigenvalues.
\end{remark}
	
\subsection{Multivariate technique of predicting FAR($p$) process}
There are multiple existing methods for the prediction of functional time series.  In Aue et al.\ (2015), a dimension-reduction method for prediction of stationary functional time series  is proposed which is demonstrated to provide competitive prediction results. To determine the order of the FAR model and the dimension of the auxiliary projected eigenspace, the functional final prediction error criterion was proposed. The FAR($p$) process is defined by the stochastic recursion
\[
Y_k-\mu=\sum_{h=1}^p\Phi_h(Y_{k-h}-\mu)+e_k,
\]
where $(e_k\colon k\in\mathbb{N})$ are centered, independent and identically distributed innovations in $L^2_H$ and  $\Phi_h \colon H \to H$ are bounded linear operators so that the above recursive equation has a unique causal solution. The multivariate prediction algorithm proceeds in three steps. 
\begin{description}
\item[Step 1.] Fix the dimension $d$, obtain the empirical fPC scores $y_{k,j}^e=\langle Y_k-\mu,\hat{v}_j\rangle $ for each observation $Y_k$, $k=1,\ldots,n$, $j=1,\ldots,d$, and the emperical $d$-variate fPC score vectors ${\bf{Y}}_k^e=(y_{k,1}^e,\ldots,y_{k,d}^e)',$ $k=1,\ldots,n$.
\item[Step 2.] Fix the order $p$, and construct a VAR model ${\bf{Y}}_{k}=\sum\limits_{j=1}^{p}\bm{\Phi}_j{\bf{Y}}_{k-j}+{\bf E}_{k}$ for the eigenscore vectors to produce the prediction $\widehat{\bf{Y}}_{n+1}^e=(\hat{y}_{n+1,1}^e,\ldots,\hat{y}_{n+1,d}^e)'.$ Durbin--Levinson and innovations algorithm can be readily applied here, given the vectors ${\bf{Y}}_1^e,\ldots,{\bf{Y}}_n^e$. 
\item[Step 3.] The predicted eigenscores are transformed back to a functional trajectory. This transformation is achieved by the truncated Karhunen-Lo\`eve representation $\widehat{Y}_{n+1}=\hat{\mu}+\hat{y}_{n+1,1}^e\hat{v}_1+\cdots+\hat{y}_{n+1,d}^e\hat{v}_d.$
\end{description}
Based on the predicted fPC scores and the estimated eigenfunctions, the resulting $\widehat{Y}_{n+1}(t)$ is then used as the $1$-step ahead prediction of $Y_{n+1}(t)$. 

\begin{remark}
This method is easy-to-implement for functional time series prediction. However, it only incorporates the correlation across trajectories and is, therefore, inferior to the PFP method when only partial trajectories are to be predicted. The optimal values of $p$ and $d$ are obtained by minimizing the final functional prediction error as developed in Aue et al.~(2015).
\end{remark}

\subsection{Fully Functional Regression Model}
\subsubsection{General setting}
\label{s2.3}
In a fully functional regression model, both the explanatory and response components are functions. Consider random explanatory functions $X(s)$ and response functions  $Y(t)$. Denote the mean functions by $\mu_X(s) = \mathrm{E}\{X(s)\}$ and $\mu_Y(t) = \mathrm{E}\{Y(t)\}$, and their covariance functions by $C_X(s_1, s_2) = \mathrm{cov}(X(s_1), X(s_2))$ \ and \ $C_Y(t_1, t_2) = \mathrm{cov}(Y(t_1), Y(t_2))$. Suppose that the Karhunen-Lo\`eve expansions of the trajectories $X$ and $Y$ are
\[
X(s)=\mu_X(s)+\sum_{i=1}^{\infty}\xi_i\phi_i(s) \qquad\mbox{and}\qquad Y(t)=\mu_Y(t)+\sum_{j=1}^{\infty}\zeta_j\psi_j(t),
\]
where $\xi_i$'s and $\phi_i$'s ($\zeta_j$'s and $\psi_j$'s) are the fPC scores and eigenfunctions of $C_X$ ($C_Y$). The fully functional linear regression model is written as
\[
Y(t)=\mu_Y(t)+\int\beta(t,s)(X(s)-\mu_X(s))ds+\tilde{e}(t),
\]
where $\tilde{e}(t)$'s are $i.i.d.$ innovation functions with mean zero, and the bivariate regression kernel $\beta(t,s)$ is assumed to be continuous and square-integrable. By the basis expansion, $\beta(t,s)$ admits the representation $\beta(t,s)=\sum\limits_{i=1}^{\infty}\sum\limits_{j=1}^{\infty}\beta_{ij}\phi_j(s)\psi_i(t).$

Without any loss of generality, the mean function of $X$ and $Y$ are now assumed to be both zero. Replacing $Y(t)$ and $X(s)$ with their Karhunen-Lo\'eve representation leads to $\sum\limits_{i=1}^{\infty} \zeta_i\psi_i(t)=\sum\limits_{i=1}^{\infty}\sum\limits_{j=1}^{\infty}\beta_{ij}\xi_j\psi_i(t)+\tilde{e}(t).$ For any $i\in\mathbb{N}_+$, taking the inner product with $\psi_i(t)$ on both sides yields $\zeta_i=\sum\limits_{j=1}^{\infty}\beta_{ij}\xi_j+u_i,$ where $u_i=\langle \tilde{e},\psi_i\rangle$. In practice, only the first  few ($d_x$) fPCs are adopted as predictors, so we consider the following equation
\begin{equation}\label{Eq:PC}
\zeta_i=\sum_{j=1}^{d_x}\beta_{ij}\xi_j+\eta_i,
\end{equation}
where $\eta_i=u_i+\sum\limits_{j>d_x}\beta_{ij}\xi_j$. Then, incorporating the first $d_y$ fPC scores of response functions resembles $d_y$ multivariate linear regression models. The estimators of $\beta_{ij}$ are obtained by fitting a regression model to the $d_y$-dimensional eigenscore vectors of the responses against the $d_x$-dimensional eigenscore vectors  of the explanatory 
functions as presented in Equation~(\ref{Eq:PC}).  Thus,  the eigenscores $\xi$'s and $\zeta$'s are first estimated and then the $\beta_{ij}$'s are estimated by fitting multiple multivariate linear regression models. As for prediction, the first step is to predict the eigenscores of $Y$ with the fitted multiple multivariate linear regression models. Then construct the predicted curve $\widehat{Y}$ by the truncated Karhunen-Lo\`{e}ve expansion $\widehat{Y}(t)=\hat{\mu}_Y(t)+\sum\limits_{j=1}^{d_y}\hat{\zeta_j}\hat{\psi}_j(t)$.

\subsubsection{Dimension selection for fully functional regression model}
\label{s2.3.2}
A common approach is to select the minimum number of principal components so that proportion of variance explained by the functional principal components exceeds a pre-specified threshold. However, the main goal here is prediction and thus it might not be always appropriate to select principal components that explain a large portion of variance.  We consider a new criterion for selecting the best dimensions of eigen-spaces of predictors and responses.  Here,  we propose to choose the dimensions by minimizing the mean squared error (MSE) of prediction, which is asymptotically equivalent to the functional final prediction error (fFPE)
\[
\mathrm{fFPE}(d_x,d_y)= \frac{n+d_x}{n}\mathrm{tr}(\widehat{\Sigma}_\eta)+\sum_{\ell>d_y}\hat{\lambda}^Y_\ell,
\]
where ${\lambda}^Y_\ell$ is the $\ell$-th eigenvalue of $C_Y$, ${\Sigma}_\eta$ is the covariance matrix of the random vector $(\eta_1,\ldots,\eta_{d_y})$, and $\widehat{\Sigma}_\eta$ is the unbiased estimator of ${\Sigma}_\eta$. The best $d_x$ and $d_y$ is the minimizer the fFPE function. The consistency of the fFPE criterion is demonstrated in Section~\ref{theory}, and the procedure is derived and explained in detail in the supplementary material.

\subsubsection{Intra-curve prediction with functional regression}	
Define $Z$ being a zero-mean random function in $L^2[0,1]$. In a regression setting for intra-curve prediction, for any $0<\tau<1$, the sub-curve over $[0, \tau]$, denoted as $Z(s)|_{[0,\tau]}=(Z(s)\colon s \in [0,\tau])$, serves as the explanatory function, and the sub-curve over $(\tau, 1]$, denoted as $Z(t)|_{(\tau,1]}=(Z(t)\colon t\in (\tau,1])$, serves as the response function. Suppose that the Karhunen-Lo\`eve expansion of the two functional variables are
\[
Z(s)|_{[0,\tau]}=\sum_{j=1}^{\infty}\xi^{(\tau)}_{j}\phi^{(\tau)}_j(s) \qquad\mbox{and}\qquad Z(t)|_{(\tau,1]}=\sum_{i=1}^{\infty}\zeta^{(\tau)}_{i}\psi^{(\tau)}_i(t),
\] 
where the notations $\xi^{(\tau)}_{j}$, $\phi^{(\tau)}_{j}(s)$, $\zeta^{(\tau)}_{i}$ and $\psi^{(\tau)}_{i}(t)$ are defined analogously to those on the entire domain $[0,1]$, but they correspond to the sub-domains $[0,\tau]$ or $(\tau,1]$. Then we consider a fully functional linear regression model
\[
Z(t)|_{(\tau,1]}=\int_0^\tau\beta_{\tau}(t,s)Z(s)|_{[0,\tau]}ds+\tilde{e}(t).
\]
Here, given any fixed value of $\tau$, assume the bivariate regression function $\beta_{\tau}(t,s)$ to be 
continuous and square integrable. Suppose now that $\beta_{\tau}(t,s)=\sum\limits_{i=1}^\infty\sum\limits_{j=1}^\infty\beta_{\tau,ij}\phi^{(\tau)}_j(s)\psi^{(\tau)}_i(t)$, then the functional regression model is expressed as  
\[
Z(t)|_{(\tau,1]}=\sum\limits_{i=1}^\infty\sum\limits_{j=1}^\infty\beta_{\tau,ij}\xi^{(\tau)}_{j}\psi^{(\tau)}_{i}(t)+\tilde{e}(t),
\]
where $\beta_{\tau,ij}$ are the unknown regression parameters that will be estimated. Under the continuity assumption on $\beta_{\tau}(t,s)$ along with $\tau$, it follows that $\beta_{\tau,ij}$ is also continuous over $\tau\in(0,1]$ for any $i$ and $j$. 
\begin{remark}
{Chiou (2012) employed this method to update prediction for traffic flow trajectories. However, as trajectories in a functional times series can be correlated, it is necessary to include appropriate steps that account for the dependence across trajectories.}
\end{remark}

\section{Partial Functional Prediction (PFP)}
\subsection{Decomposition of functional time series}
As has been discussed in the introduction, an well-established model for smooth functional time series is referred to the following decomposition framework,
\[
Y_k(t)=S_k(t)+\epsilon_k(t),\hspace{0.7cm} t\in [0,1],
\]
where {$S_k(t)$ is the signal correlated to the previous trajectories and independent with the future innovations by causality}, and $\epsilon_k(t)$ is the innovation function that is independent with the previous trajectories. If the observed trajectories are contaminated by random noise, we decompose the functional time series into three parts:
\[
Y_k(t_j)=S_k(t_j)+\epsilon_k(t_j)+e_k(t_j),\hspace{7mm} k=1,\ldots,n,\\\ j\ge1.
\]
Note that $\epsilon_k(t_j)$ and $e_k(t_j)$ have different roles in the sense that, $e_k(t_j)$ represents random error accounting for the short-term {pointwise} dynamics and the roughness of functions, while $\epsilon_k(t_j)$ represents functional innovation accounting for the long-term functional dynamics. In practice, the observations are available only at pre-specified discrete grids, so here we use $t_j$ instead of $t$. 
We propose a step-wise procedure to predict all components. The predicted components are combined to form the final prediction. 
	
\subsection{Smooth functions}
Suppose that we have observed $Y_1,\ldots,Y_n$, and $Y_{n+1}|_{[0,\tau]}$. The updated prediction of the curve over $(\tau,1]$ is given by $\widehat{Y}^u_{n+1}|_{(\tau,1]}=\widehat{Y}_{n+1}|_{(\tau,1]}+\hat{\epsilon}_{n+1}|_{(\tau,1]},$ where $\widehat{Y}_{n+1}$ is the ``next-interval'' prediction of $Y_{n+1}$ and $\hat{\epsilon}_{n+1}|_{(\tau,1]}$ is the intra-curve prediction of the $(n+1)$-th innovation function over $(\tau,1]$.

To predict ${\epsilon}_{n+1}|_{(\tau,1]}$, we consider a fully functional regression model, where $(\epsilon_{k}(s)|_{[0,\tau]}\colon k\in\mathbb{N})$ serve as the predictors and $(\epsilon_{k}(t)|_{(\tau,1]}\colon k\in\mathbb{N})$ serve as the responses,
\[
\epsilon_{k}(t)|_{(\tau,1]}=\int_0^\tau\beta_{\tau}(t,s)\epsilon_{k}(s)|_{[0,\tau]}ds+\tilde{e}_k(t).
\]
By the Karhunen-Lo\`eve expansion, 
\[
\epsilon_{k}(s)|_{[0,\tau]}=\sum_{j=1}^{\infty}\xi_{kj}^{(\tau)}\phi^{(\tau)}_j(s)\qquad\mbox{and}\qquad \epsilon_{k}(t)|_{(\tau,1]}=\sum_{i=1}^{\infty}\zeta^{(\tau)}_{ki}\psi^{(\tau)}_i(t).
\]
The innovation function is unobserved, so the functional regression model is applied to the prediction residuals $\tilde{\epsilon}_k=Y_k-\widehat{Y}_k$ ($k\le n$) of the previous functions. Replacing the unknown terms with the estimated (predicted) values and adopting the first $d_x$ and $d_y$ fPCs for predictors and responses respectively lead to the prediction of ${\epsilon}_{n+1}(t)|_{(\tau,1]}$ as follows
${\hat{\epsilon}}_{n+1}(t)|_{(\tau,1]}=\sum\limits_{i=1}^{d_y}\sum\limits_{j=1}^{d_x}\hat{\beta}_{\tau,ij}\hat{\xi}^{(\tau)}_{n+1,j}\hat{\psi}^{(\tau)}_i(t),$ and the final prediction of $Y_{n+1}|_{(\tau,1]}$ is $\widehat{Y}^u_{n+1}|_{(\tau,1]}=\widehat{Y}_{n+1}|_{(\tau,1]}+{\hat{\epsilon}}_{n+1}|_{(\tau,1]}.$

The updated prediction $\widehat{Y}^u_{n+1}|_{(\tau,1]}$ can be regarded as the complete curve prediction $\widehat{Y}_{n+1}|_{(\tau,1]}$ adjusted by the intra-curve prediction of the $(\tau,1]$ block of the residual function ${\hat{\epsilon}}_{n+1}|_{(\tau,1]}$. The prediction steps are now summarized by the following algorithm.
\begin{description}
\item[Step 1.] Fix $d$ and $p$, apply ``next-interval'' prediction to obtain the prediction of the entire curve $\widehat{Y}_{n+1}$ and the prediction residuals $(\tilde{\epsilon}_k(t)\colon k=p+1,\ldots,n)$, $\tilde{\epsilon}_{n+1}(t)\vert_{[0,\tau]}$.
\item[Step 2.] Segment the prediction residual functions in Step 1 at ``current time'' $\tau$. Treat the first parts $(\tilde{\epsilon}_{k}|_{[0,\tau]})_{k=p+1}^{n}$ as the predictors, and the second parts $(\tilde{\epsilon}_{k}|_{(\tau,1]})_{k=p+1}^{n}$ as the responses. Fix $d_x$ and $d_y$, and apply intra-curve functional regression of $(\tilde{\epsilon}_{k}|_{(\tau,1]})_{k=p+1}^{n}$ on $(\tilde{\epsilon}_{k}|_{[0,\tau]})_{k=p+1}^{n}$, and use the fitted model to obtain the prediction of the $(\tau,1]$ block of the $(n+1)$-th innovation function ${\hat{\epsilon}}_{n+1}|_{(\tau,1]}$.
\item[Step 3.] Add the $(\tau,1]$ segment of the complete predicted curve  $\widehat{Y}_{n+1}$ and ${\hat{\epsilon}}_{n+1}|_{(\tau,1]}$ to get the final prediction $\widehat{Y}^u_{n+1}|_{(\tau,1]}=\widehat{Y}_{n+1}|_{(\tau,1]}+{\hat{\epsilon}}_{n+1}|_{(\tau,1]}$.
\end{description}

\subsection{Noisy functions}
In this section, we consider functional data as noisy sampled points from a collection of consecutive trajectories. In practice, the observed functional time series is observed at discrete time grids, thus the observed trajectories can be rough.  This may be due to measurement errors or sparsely-spaced observation time grids. As discussed in Yao et al.\ (2005), the random error leads to biased fPC scores. To prevent this problem, the raw trajectories are preprocessed by smoothing. However, in the random error $(e_k(t_j),k\in\mathbb{N},~j \ge1)$, which is not smooth, there can still exist short-term temporal dependence, {thus additional steps are necessary to utilize the dependence information in $(e_k(t_j),k\in\mathbb{N},~j \ge1)$} to improve the prediction.
	
As previously discussed, any noisy functional time series $(Y_k(t),k\in\mathbb{N})$ can be decomposed into three parts,
\[
Y_k(t_j)=S_k(t_j)+\epsilon_k(t_j)+e_k(t_j),\hspace{7mm} k\in\mathbb{N},\ j\ge1,
\]
where $S_k(t_j)$ is the smooth signal depending on the smooth part of the past functions, $\epsilon_k(t_j)$ is the independent smooth innovation function, and $e_k(t_j)$ is the random error. Let $f_k(t_j)=S_k(t_j)+\epsilon_k(t_j)$ represent the smooth part of the functional time series, which can be estimated by smoothing techniques (e.g.~kernel smoothing, basis smoothing, local polynomial regression). Now fit the ARMA model to the smoothing residuals $(r_k(t_j))_{k,j\ge1}$, defined as $r_k(t_j)=Y_k(t_j)-\tilde{f}_k(t_j)$, where $(\tilde{f}_k(t)\colon k\in\mathbb{N})$ are the smoothed trajectories. For noisy trajectories, there are additional steps in the algorithm.

\begin{description}
\item[Step 4.] Produce the smoothed curve $(\tilde{f}_k(t)\colon k\in\mathbb{N})$ with any smoothing technique and denote the smoothing residuals to be $(r_k(t_j))_{k,j\ge1}$. Apply ARMA model to the smoothing residuals to predict the future random errors $\hat{r}_{n+1}(t_j)$, $t_j>\tau$.
\item[Step 5.] Combine the prediction of the smooth part in Step 3, $\hat{f}_{n+1}(t)$, and the prediction of the smoothing residuals to obtain the final prediction $\widehat{Y}^u_{n+1}(t_j)=\hat{f}_{n+1}(t_j)+\hat{r}_{n+1}(t_j).$
\end{description}	 

This adjustment is necessary if the observed trajectories are significantly rough and the dependence of $(e_k(t_j))_{k,j\ge1}$ across $t_j$ is pronounced. The auto-correlation of the smoothing residuals typically decays much faster than that of the original time series, which indicates that the long-term dynamics (e.g.\ seasonal trend) is removed.

\subsection{Convergence rate and criterion consistency}	
\label{theory}
In this section, we develop the upper bound of the convergence rate of the partial prediction and the consistency of the fFPE criterion. The convergence rate is related to the fPC scores of the current partial residual $\tilde{\epsilon}_{n+1}(t)\vert_{[0,\tau]}$, the correlation between the two segments of prediction residuals, the decay rates of the eigenvalues of the covariance function of the first-part residual $\tilde{\epsilon}_k(t)\vert_{[0,\tau]}$ and the second-part residual $\tilde{\epsilon}_k(t)\vert_{(\tau,1]}$, where $\tilde{\epsilon}_{k}(t)=Y_{k}-\mu-\sum\limits_{h=1}^p\Phi_{h,d}(Y_{k-h,d}-\mu),$ and $\Phi_{h,d}$, $Y_{k,d}$ are the $d$-dimensional projection of $\Phi_{h}$, $Y_{k}$ on $\{v_{j}\colon j=1,\ldots,d\}$. Define
$$\mathcal{G}_{\tau}(t,s)=\text{cov}\{\tilde{\epsilon}_k(t)\vert_{(\tau,1]},\tilde{\epsilon}_k(s)\vert_{[0,\tau]}\},$$
$$C_{\epsilon,1}(t,s)=\text{cov}\left\{\tilde{\epsilon}_k(t)\vert_{[0,\tau]},\tilde{\epsilon}_k(s)\vert_{[0,\tau]}\right\},\qquad C_{\epsilon,2}(t,s)=\text{cov}\left\{\tilde{\epsilon}_k(t)\vert_{(\tau,1]},\tilde{\epsilon}_k(s)\vert_{(\tau,1]}\right\},$$
and the estimators are
$$\widehat{\mathcal{G}}_{\tau}(t,s)=\frac{1}{n}\sum_{k=1}^n\tilde{\epsilon}_k(t)\vert_{(\tau,1]}\tilde{\epsilon}_k(s)\vert_{[0,\tau]},$$
$$\widehat{C}_{\epsilon,1}(t,s)=\frac{1}{n}\sum_{k=1}^n\tilde{\epsilon}_k(t)\vert_{[0,\tau]}\tilde{\epsilon}_k(s)\vert_{[0,\tau]},\qquad \widehat{C}_{\epsilon,2}(t,s)=\frac{1}{n}\sum_{k=1}^n\tilde{\epsilon}_k(t)\vert_{(\tau,1]}\tilde{\epsilon}_k(s)\vert_{(\tau,1]}.$$
By the Mercer's theorem, suppose that $C_{\epsilon,1}(t,s)$ and $C_{\epsilon,2}(t,s)$ admit the following spectral decomposition 
$$C_{\epsilon,1}(t,s)=\sum_{j=1}^\infty\theta_j\phi_{j}(t)\phi_{j}(s),\qquad C_{\epsilon,2}(t,s)=\sum_{i=1}^\infty\omega_i\psi_{i}(t)\psi_{i}(s).$$
Consider the intra-curve regression model $\tilde{\epsilon}_k(t)\vert_{(\tau,1]}=\int\beta_\tau(t,s)\tilde{\epsilon}_k(s)\vert_{[0,\tau]}ds+\delta_k(t)$ and represent $\beta_\tau(t,s)$ with $\phi_j(s)$'s and $\psi_i(t)$'s as $\beta_\tau(t,s)=\sum\limits_{i,j=1}^{\infty}\beta_{\tau,ij}\psi_i(t)\phi_{j}(s).$
Notationally let
$\bm{V}_1(s)=(\phi_{1}(s),\ldots,\phi_{d_x}(s)), \bm{V}_2(t)=(\psi_{1}(t),\ldots,\psi_{d_y}(t))$ and
$$G=\int_\tau^1\int_0^\tau \bm{V}^T_2(t)\mathcal{G}_{\tau}(t,s)\bm{V}_1(s)dsdt,\qquad \widehat{G}=\int_\tau^1\int_0^\tau \widehat{\bm{V}}^T_2(t)\widehat{\mathcal{G}}_{\tau}(t,s)\widehat{\bm{V}}_1(s)dsdt,$$
\begin{align*}
B_{11}=\left(
\begin{tabular}{ccc}
$\beta_{\tau,11}$&$\cdots$&$\beta_{\tau,1d_x}$\\
$\vdots$&$\ddots$&$\vdots$\\
$\beta_{\tau,d_y1}$&$\cdots$&$\beta_{\tau,d_yd_x}$
\end{tabular}\right),&\qquad
\Theta=\left(
\begin{tabular}{ccc}
$\theta_1$&&\\
&$\ddots$&\\
&&$\theta_{d_x}$
\end{tabular}\right).
\end{align*}
Note that $G$ and $\widehat{G}$ are $d_y\times d_x$ matrices. A truncated estimator of the unknown coefficient $\beta_\tau(t,s)$ can be represented as $\hat{\beta}_\tau(t,s)=\widehat{\bm{V}}_2(t)\widehat{B}_{11}\widehat{\bm{V}}^{T}_1(s),$ where 
$$\widehat{\bm{V}}_1(s)=(\hat{\phi}_{1}(s),\ldots,\hat{\phi}_{d_x}(s)),\qquad \widehat{\bm{V}}_2(t)=(\hat{\psi}_{1}(t),\ldots,\hat{\psi}_{d_y}(t)),$$ and $\widehat{B}_{11}=\widehat{G}\widehat{\Theta}^{-1}.$ We represent $\tilde{\epsilon}_{n+1}(t)\vert_{[0,\tau]}$ as $\tilde{\epsilon}_{n+1}(t)\vert_{[0,\tau]}=\sum\limits_{j\ge1}x_j\phi_{j}(t)$, where $x_j=\langle \tilde{\epsilon}_{n+1}\vert_{[0,\tau]},\phi_j\rangle$. Here $\tilde{\epsilon}_{n+1}(t)\vert_{[0,\tau]}$ is treated as a fixed function. The following assumptions are made.
\begin{itemize}
\item[(A.3)] With some $\alpha,\alpha_1,\alpha_2>1$, $C_y,C_1,C_2>0$, the eigenvalue sequences $(\lambda_{i}\colon i\in\mathbb{N})$, $(\theta_{i}\colon i\in\mathbb{N})$ and $(\omega_{i}\colon i\in\mathbb{N})$ are strictly decreasing, and satisfy the conditions below
\begin{align*}
&C_y^{-1}i^{-\alpha}\le\lambda_{i}\le C_{y}i^{-\alpha},\qquad C_1^{-1}i^{-\alpha_1}\le\theta_{i}\le C_1i^{-\alpha_1},\qquad C_2^{-1}i^{-\alpha_2}\le\omega_{i}\le C_2i^{-\alpha_2},\\
&\hspace{2cm}\delta_{1i}\equiv\theta_i-\theta_{i+1}\ge C_{1}i^{-\alpha_1-1},\qquad \delta_{2i}\equiv\omega_i-\omega_{i+1}\ge C_2i^{-\alpha_2-1}.
\end{align*}
\item[(A.4)] There exists some positive constants $\beta_1$, $\beta_2$, and $\gamma$, so that $\abs{G_{ij}}\equiv\abs{\langle\int\mathcal{G}_{\tau}\phi_j,\psi_i\rangle} \le \mbox{const.}i^{-\beta_2}j^{-\beta_1}$ and $\abs{x_j}\le \mbox{const.}j^{-\gamma},$
where ``$\mbox{const.}$'' denotes a constant.
\item[(A.5)] $\gamma>1/2$, $\alpha_1-\beta_1<-\frac{1}{2}$, $\alpha_2-\beta_2<-\frac{1}{2}$, $\alpha_1>\max\{1,(2\gamma-3)/4\}$, $\alpha_2>1$.
\item[(A.6)] Define $\mathcal{Y}_k(t,s)=\tilde{\epsilon}_k(t)\tilde{\epsilon}_k(s)-E\{\tilde{\epsilon}_k(t)\tilde{\epsilon}_k(s)\}$, and we assume that $p$ and $d$ are selected so that
\begin{align*}
\sum_{\substack{1\le\abs{r_1}<n}}\sum_{\substack{1\le i\le n-\abs{r_1}}}&\int\bigg|E\bigg\{\mathcal{Y}_i(t_1,s_1)\mathcal{Y}_{i+r_1}(t_1,s_1)\bigg\}\bigg|dt_1ds_1=O(n),\\
\sum_{\substack{1\le\abs{r_1}<n\\1\le\abs{r_2}<n}}\sum_{\substack{1\le i\le n-\abs{r_1}\\ 1\le j\le n-\abs{r_2}}}&\int\int\bigg|E\bigg\{\mathcal{Y}_i(t_1,s_1)\mathcal{Y}_{i+r_1}(t_1,s_1)\\
&\hspace{1.7cm}\mathcal{Y}_j(t_2,s_2)\mathcal{Y}_{j+r_2}(t_2,s_2)\bigg\}\bigg|dt_1ds_1dt_2ds_2=O(n^2),\\
\sum_{\substack{1\le\abs{r_1}<n\\1\le\abs{r_2}<n\\1\le\abs{r_3}<n}}\sum_{\substack{1\le i\le n-\abs{r_1}\\ 1\le j\le n-\abs{r_2}\\ 1\le k\le n-\abs{r_3}}}&\int\int\int\bigg|E\bigg\{\mathcal{Y}_i(t_1,s_1)\mathcal{Y}_{i+r_1}(t_1,s_1)\mathcal{Y}_j(t_2,s_2)\\
&\mathcal{Y}_{j+r_2}(t_2,s_2)\mathcal{Y}_k(t_3,s_3)\mathcal{Y}_{k+r_3}(t_3,s_3)\bigg\}\bigg|dt_1ds_1dt_2ds_2dt_3ds_3=O(n^3).
\end{align*}
\item[(A.7)] Each $Y_k$ admits the representation $Y_k=f(u_k,u_{k-1},\ldots),$ where $(u_k\colon k\in\mathbb{N})$ are $i.i.d$ elements taking values in a measurable space $S$, and $f$ is a measurable function $f\colon S^\infty\to H$. Moreover, we assume that if $(u'_k\colon k\in\mathbb{N})$ is an independent copy of $(u_k\colon k\in\mathbb{N})$ defined on the same probability space, then letting $Y_k^{(m)}=f(u_k,u_{k-1},\ldots,u_{k-m+1},u'_{k-m},u'_{k-m-1},\ldots),$
we have $\sum\limits_{k=1}^{\infty}(E\|Y_k-Y_k^{(k)}\|_4^4)^{1/4}<\infty$, where $\|\cdot\|_4$ signifies the $\ell_4$-norm.
\end{itemize}

From Assumption (A.3) and (A.4), it follows that $\abs{\beta_{\tau,ij}}\le \mbox{const.}i^{-\beta_2}j^{\alpha_1-\beta_1}.$ Assumption (A.3) assures the identifiability of principal components and consistency of estimation (see e.g.\ Gohberg, I.\ and Krupnik, N., 1992). Assumption (A.4) quantifies the correlation between the two parts of residual functions. Assumption (A.5) assures the square integrability of $\tilde{\epsilon}_{n+1}\vert_{[0,\tau]}$ and $\beta_\tau(t,s)$, and $\alpha_2-\beta_2<-1/2$ is an analogical assumption of $\alpha_1-\beta_1<-1/2$, which indicates that if the predictor and response are switched, the intra-curve regression model is still well defined. $\alpha_1>(2\gamma-3)/4$ restricts that the fPC scores of the current partial residual does not decay much faster than average, and the assumptions $\alpha_1,\alpha_2>1,\gamma>1/2$ come from the assumption $Y_k\in L^2[0,1]$. Assumption (A.6) and (A.7) assure that the weak dependence across the prediction residual functions does not influence the convergence rate, which are naturally true for $m$-dependent processes.

The upper bound of the convergence rate of the partial prediction is presented in the following theorem. 
\begin{theorem}
\label{th3}
Set $d_x=O(n^{1/\tau_1})$ and $d_y=O(n^{1/\tau_2})$, where 
$\tau_1>2(\alpha_1+1)$ and $\tau_2>\max\{2(\alpha_2+1),(2\alpha_2+3)\tau_1/(2\tau_1-1)\}$. 
Under Assumptions (A.3)---(A.7), we can find $\alpha_1$ and $\alpha_2$ (where $\alpha_1,\alpha_2$ 
are related to $\tau$), such that $\alpha_1>\alpha$ and $\alpha_2>\alpha$ for any $\tau\in(0,1)$. Let $\rho=\max\{\frac{2(\alpha_1-\beta_1-\gamma+1)}{\tau_1}\vee (\frac{-2\beta_2+1}{\tau_2})\vee (\frac{1}{\tau_1}+\frac{2\alpha_2+3}{\tau_2}-2)\vee (\frac{4\alpha_1-2\gamma+4}{\tau_1}-2)\}$, then
\begin{align*}
&E\left\|\int(\hat{\beta}_\tau-\beta_\tau)\tilde{\epsilon}_{n+1}\vert_{[0,\tau]}\right\|^2\le\mbox{const.}n^\rho\\
&\vee\left\{
\begin{array}{rcl}
n^{\{(\frac{\alpha_1-2\gamma+2}{\tau_1}-1)\vee(\frac{4\alpha_1-2\beta_1-2\gamma+2}{\tau_1}-1)\}}, & & \mbox{if}\ {\alpha_1-2\gamma>-1,2\alpha_1-\beta_1-\gamma>-1}\\
n^{\frac{1}{\tau_1}-1}\log(n)\vee n^{\frac{4\alpha_1-2\beta_1-2\gamma+2}{\tau_1}-1}, & & \mbox{if}\ {\alpha_1-2\gamma=-1,2\alpha_1-\beta_1-\gamma>-1}\\
n^{\{(\frac{1}{\tau_1}-1)\vee(\frac{4\alpha_1-2\beta_1-2\gamma+2}{\tau_1}-1)\}}, & & \mbox{if}\ {\alpha_1-2\gamma<-1,2\alpha_1-\beta_1-\gamma>-1}\\
n^{\frac{\alpha_1-2\gamma+2}{\tau_1}-1}, & & \mbox{if}\ {\alpha_1-2\gamma>-1,2\alpha_1-\beta_1-\gamma\le-1}\\
n^{\frac{1}{\tau_1}-1}\log(n), & & \mbox{if}\ {\alpha_1-2\gamma=-1,2\alpha_1-\beta_1-\gamma\le-1}\\
n^{\frac{1}{\tau_1}-1}, & & \mbox{if}\ {\alpha_1-2\gamma<-1,2\alpha_1-\beta_1-\gamma\le-1}
\end{array} \right..
\end{align*}
\end{theorem}
\begin{remark}
In Theorem 1, ``$a\vee b$'' signifies $\max\{a,b\}$. A practical implication of $\alpha_1>\alpha$ and $\alpha_2>\alpha$ is that, compared to the entire function, partial trajectories can be approximated with less principal components, while equal proportion of variance is explained. The assumptions on $\tau_1$ and $\tau_2$ guarantee the convergence of prediction and uniform consistency. Otherwise, if $d_x$ and $d_y$ were selected too large, the prediction would diverge. 
\end{remark}
The following theorem demonstrates the consistency of the fFPE criterion in Section~\ref{s2.3.2}. Define $\Delta=n^{-1}\sum\limits_{k=1}^n\tilde{\epsilon}_k(t)\tilde{\epsilon}_k(s)-E\{\tilde{\epsilon}_k(t)\tilde{\epsilon}_k(s)\}$, then if the following assumptions hold, we can develop the consistency property of the fFPE criterion.
\begin{itemize}
\item[(A.8)] $E\|\Delta\|^{16}=O(n^{-8})$, $n^{-1}d^{3\alpha_2+2}_y\to0$, $n^{-1}d_y\left(\sum\limits_{i=1}^{d_y}\delta^{-4}_{2i}\right)^{1/2}\to0$, and
$$n^{-1}d_x\sum_{i=1}^{d_y}\sum_{j=1}^{d_x}\{G^2_{ij}\theta_j^{-2}(\delta_{1j}^{-2}+\theta_j^{-1})+\omega_i\theta_j^{-1}\delta_{1j}^{-2}+\delta_{2i}^{-2}\}\to0.$$
\end{itemize}
\begin{theorem}	
Let $\hat{\epsilon}^u_{n+1}\vert_{(\tau,1]}$ be the prediction of $\tilde{\epsilon}_{n+1}\vert_{(\tau,1]}$ based on $C_{\epsilon,1}$ and $C_{\epsilon,2}$, and $\tilde{\epsilon}^u_{n+1}\vert_{(\tau,1]}$ be the prediction of $\tilde{\epsilon}_{n+1}\vert_{(\tau,1]}$ based on $\widehat{C}_{\epsilon,1}$ and $\widehat{C}_{\epsilon,2}$. Then, under Assumptions (A.3), (A.6) and (A.8), we have
\[
E[\|\tilde{\epsilon}_{n+1}\vert_{(\tau,1]}-\hat{\epsilon}^u_{n+1}\vert_{(\tau,1]}\|^2]-E[\|\tilde{\epsilon}_{n+1}\vert_{(\tau,1]}-\tilde{\epsilon}^u_{n+1}\vert_{(\tau,1]}\|^2]\to 0
\] 
uniformly for $d_x$ and $d_y$ as $n\to\infty.$
\end{theorem}
\begin{remark}
In practice, $\Phi_{h}(\cdot)$'s are unknown and need to be estimated. However, if Assumptions (A.3)---(A.8) still hold for the estimated residuals, the result still follows.
\end{remark}

\section{Simulation}
\subsection{General setting}
To analyze the finite sample properties of PFP, a comparative simulation study 
was conducted. The PFP method was tested on simulated FAR models. In each simulation test, 400 trajectories were generated by a FAR($p$) model ($p=1$ or $2$). The $(\tau,1]$ blocks of the last 20 trajectories were predicted. The corresponding mean squared error (MSE) of prediction was computed, 
as well as the fFPE value for comparison. This procedure was repeated for 100 times for each simulation setup.
	
The simulation was conducted under the context of a $D$-dimensional functional space which is spanned by the Fourier basis functions $\bm{\nu}=(\nu_1,\nu_2,\ldots,\nu_D)$ on the unit interval $[0,1]$ that correspond to the first $(D-1)/2$ fundamental Fourier frequencies. Thus, any simulated function admits the representation $x(t)=\sum\limits_{j=1}^{D}c_j\nu_j(t)$ with coefficients $\bm{c}=(c_1,\ldots,c_D)'$. Then for any linear 
operator $\Psi\colon H\to H$, 
\[
\Psi(x)=\sum_{j=1}^Dc_j\Psi(\nu_j)=\sum_{j=1}^D\sum_{j'=1}^Dc_j\langle \Psi(\nu_j),\nu_{j'}\rangle \nu_{j'}=\bm{c'\Psi}  {\bm{\nu}},
\] 
where ${\bf \Psi}$ is a $D\times D$ matrix with elements $\{\langle \Psi(\nu_j),\nu_{j'}\rangle\}_{j,j'=1}^D$. The innovation function admits the basis expansion $\epsilon_k(t)=\sum\limits_{j=1}^D a_{k,j}\nu_j(t)$, where $a_{k,j}$'s are $i.i.d.$ normal random variables with mean zero and standard deviation $\sigma_j$. Two sets of standard deviations used here are ${\bf \sigma}_1=(j^{-1}\colon j =1,\ldots,D)$ and ${\bf \sigma}_2=(1.2^{-j}\colon j=1,\ldots,D)$.
	
\subsection{Prediction comparison for smooth trajectories}
In this section, we show the comparison of partial functional prediction with Aue et al.\ (2015)'s method and the intra-curve functional regression method on FAR(2) processes $Y_k=\Psi_1(Y_{k-1})+\Psi_2(Y_{k-2})+\epsilon_k$. The operators were generated such that $\Psi_1 = \kappa_1\Psi$ and $\Psi_2=\kappa_2\Psi$. (Here, note that $\kappa_2=0$ yields a FAR(1) process). The operator matrix ${\bf \Psi}$ was generated at random, with each element following a normal distribution with mean zero and variance $\sigma_{jj'}$, and then scaled by its $l_2$-norm. We set $\sigma_{jj'}$ to be $({\bf\sigma}_i{\bf\sigma}_i')_{jj'}$ to ensure the simulated functions satisfying Riemann-Lebesgue Lemma, and set $D=15$.
	
In each simulation setup, the MSEs of prediction $ \int_{\tau}^1\{Y_{n+1}(t)-\widehat{Y}^u_{n+1}(t)|_{(\tau,1]}\}^2dt$ of PFP and the two competitor methods were computed. The fFPE values were also calculated for PFP and the intra-curve regression, which were recorded to be close to the corresponding MSE of prediction. Results for five pairs of values $(\kappa_1,\kappa_2)$ are provided in Table 1.  For clarity,  the following abbreviations are introduced: PMSE=``prediction mean squared error'', fFPE=``final functional prediction error''. The subscript: $ts$=``FAR model prediction (Aue et al.~(2015))'', $r$=``intra-curve regression prediction''.   

\begin{table}[h!]
	\centering
	\begin{small}
	\label{my-label}
	\begin{tabular}{p{0.2in}p{0.2in}|p{0.6in}p{0.6in}p{0.6in}p{0.6in}p{0.6in}}
	\hline
	\hline
	 \multicolumn{7}{c}{\textit{$\sigma_1$}} \\
	\hline
         $\kappa_1$ & $\kappa_2$ & fFPE$_{PFP}$& PMSE$_{PFP}$& PMSE$_{ts}$& fFPE$_{r}$& PMSE$_{r}$\\
         \hline
         1.8& 0.0&0.2024& 0.2097 &0.8442 &0.3269 &0.3431 \\
         0.8& 0.0& 0.2003& 0.2112& 0.8396 &0.2664& 0.2763 \\
         0.2& 0.0& 0.1928& 0.2018& 0.8286 &0.1938& 0.2025\\
         0.4& 0.4& 0.2038& 0.2123& 0.8388 &0.2309& 0.2392\\
         0.0& 0.8& 0.2058& 0.2115& 0.8419 &0.2647& 0.2685\\
        \hline 
        \multicolumn{7}{c}{\textit{$\sigma_2$}} \\
         \hline
          $\kappa_1$ & $\kappa_2$ & fFPE$_{PFP}$& PMSE$_{PFP}$& PMSE$_{ts}$ & fFPE$_{r}$& PMSE$_{r}$\\
          \hline
          1.8& 0.0& 0.5554& 0.5801 &1.2269 &1.1012 &1.1668 \\
          0.8& 0.0&  0.5455& 0.5640& 1.2112 &0.7011 &0.7431 \\
          0.2& 0.0&  0.5302& 0.5561& 1.1813 &0.5287 &0.5536 \\
          0.4& 0.4& 0.5711 &0.5985 &1.2593 &0.6128 &0.6391 \\
          0.0& 0.8&0.5740 &0.5907& 1.2631 &0.6995& 0.7127 \\
        \hline
          \hline
\end{tabular}
\caption{fFPE and PMSE values of the three methods under different simulation setups.}
\end{small}
\end{table}

Clearly, when the signal-noise ratio is high, PFP outperforms the other methods. Otherwise, the performances of PFP and the intra-curve regression method are similar, and are both better than full-curve prediction. The fFPE value and the prediction MSE are always very close for different situations, which numerically justifies the practical applicability of the fFPE criterion. The results of bootstrap prediction interval are included in the supplementary material.
	
\subsection{Prediction comparison for noisy trajectories}
Noisy functional time series were simulated by adding errors following AR($1$) process to the smooth functional time series. Specifically, the simulated functions admit the decomposition $Y_k(t_j)=S_k(t_j)+e_k(t_j),\ j=1,\ldots,48,$ where $S_k(t_j)$ is the smooth curve obtained from the simulated FAR($1$) process ($\kappa_1=1.8$, $\kappa_2=0$), and $e_k(t_j)$ is the random error. The ``current time'' $\tau=0.5$. 

\afterpage{	
\doublespacing
	\begin{landscape}
	\centering

	\label{tab2}
	\begin{tabular}{|p{0.4in}|p{0.41in}p{0.41in}p{0.41in}p{0.43in}|p{0.41in}p{0.41in}p{0.41in}p{0.43in}|p{0.41in}p{0.41in}p{0.41in}p{0.43in}|}
	\hline
	\hline
	\multicolumn{1}{|c|}{}&\multicolumn{4}{c|}{\textit{$\phi=0.5$, $\sigma=0.2$}}  &\multicolumn{4}{c|}{\textit{$\phi=0.5$, $\sigma=0.5$}} & \multicolumn{4}{c|}{\textit{$\phi=0.8$, $\sigma=0.5$}}\\
	\hline
        $h$ & $\mbox{PMSE}_n$ & $\mbox{PMSE}_s$ &$\mbox{PMSE}_a$ &$\mbox{PMSE}_i$ & $\mbox{PMSE}_n$ & $\mbox{PMSE}_s$ &$\mbox{PMSE}_a$ &$\mbox{PMSE}_i$ &$\mbox{PMSE}_n$ & $\mbox{PMSE}_s$ &$\mbox{PMSE}_a$ &$\mbox{PMSE}_i$ \\
         \hline
         $h=1$& 0.2692& 0.8986 &0.3965& 0.2950& 0.4917& 0.9362& 0.6295& 0.6665& 0.4657& 0.9661& 0.5919& 0.6468\\
	       &  (0.359)& (0.083)& (0.231)& (0.327)& (0.325)& (0.190)& (0.245)& (0.240) & (0.350)& (0.150)& (0.260)& (0.240) \\
	 \hline     
 	 $h=2$& 0.3711& 0.6648& 0.8446& 0.4572& 0.5365& 0.6777& 0.9872& 0.8138& 0.5307& 0.7050& 0.9745& 0.8612\\
               &  (0.392)& (0.110)& (0.159)& (0.339) & (0.355)& (0.225)& (0.175)& (0.245) & (0.425)& (0.170)& (0.170)& (0.235) \\
	 \hline
	 $h=3$&0.3297 &0.3159 &1.3165 &0.6106& 0.4715& 0.4275& 1.3991& 0.9037& 0.4819& 0.4280& 1.4347& 1.0113\\
 	       & (0.338)& (0.363)& (0.080)& (0.219)	& (0.320)& (0.395)& (0.110)& (0.175) & (0.370)& (0.370)& (0.095)& (0.165)\\
	 \hline
	 $h=4$&0.2189 &0.4064 &1.8303 &0.7634& 0.4083& 0.4678& 1.8587& 0.9607& 0.3645& 0.4560& 1.8708& 1.0901\\
 	       & (0.565)& (0.246)& (0.043)& (0.146) & (0.450)& (0.310)& (0.060)& (0.180) & (0.455)& (0.340)& (0.055)& (0.150)\\ 
	 \hline
	 $h=5$& 0.1560& 0.7263& 2.4103& 0.9459& 0.4065& 0.7583& 2.3927& 1.1331& 0.3886& 0.7587& 2.4336& 1.2443\\
	       & (0.733)& (0.130)& (0.037)& (0.100) & (0.540)& (0.210)& (0.060)& (0.190) & (0.545)& (0.235)& (0.060)& (0.160)\\ 
         \hline
         \hline
\end{tabular}
\captionof{table}{Average prediction mean squared error, and the proportion of cases where the corresponding 
prediction MSE is the minimal (in the parenthesis). $\mbox{PMSE}_n$ is the prediction MSE of PFP in the noisy case, $\mbox{PMSE}_s$ is the prediction MSE of PFP in the smooth case, $\mbox{PMSE}_a$ is the prediction MSE of ARIMA model, 
{$\mbox{PMSE}_i$ is the prediction MSE of PFP as smoothing the functions by linear interpolation.} The parameter $\phi$ is the coefficient of the AR(1) process 
of the random errors and $\sigma^2$ is the variance of the random error. The prediction procedure was 
repeated 200 times under each setting.}

	\end{landscape}
}
	
To incorporate the random errors into prediction, two other approaches were also considered: 1).\ ARIMA model, and 2).\ applying only Step 1--3 to the original time series without smoothing (or equivalently, smoothing the functions by linear interpolation).  The average prediction MSE of the following 5 grids ($1\le h\le 5$) of the last 20 trajectories are shown in Table~2. The simulation experiments indicate that the ARIMA model should be the ``last-resort" method for long-term prediction. The ARIMA model may provide reasonable short term prediction, and one can use this approach to predict the rough errors. However, if the error term is incorporated into PFP in the smooth case by linear interpolation, the prediction will deteriorate. This is because that the estimation of the actual fPC scores is biased, and this error propagates to the estimated FAR model. 
	
\section{Analysis of PM10 and Traffic flow trajectories}
\subsection{Analysis of PM10 concentration}
One goal of this section is to analyze the trajectories of PM10 concentration which broadly refers to particulate matter with an aerodynamic diameter of less than 10$\mu m$ in ambient air. The trajectories were measured every 30 minutes in Graz, Austria.  Prior to applying the proposed PFP prediction method, the data was segmented  according to the day of the week. The 48 observations for each day were combined into a vector. Visual inspection of the data revealed several extreme outliers around New Year's Eve known to be caused by firework activities. The corresponding week was removed from the sample. To stabilize the variance,  the square root transformation was applied to the recordings. The remaining 175 transformed trajectories are 
displayed in Figure 2. 
	\begin{figure}[!h]
	\centering
	\includegraphics[scale=0.45]{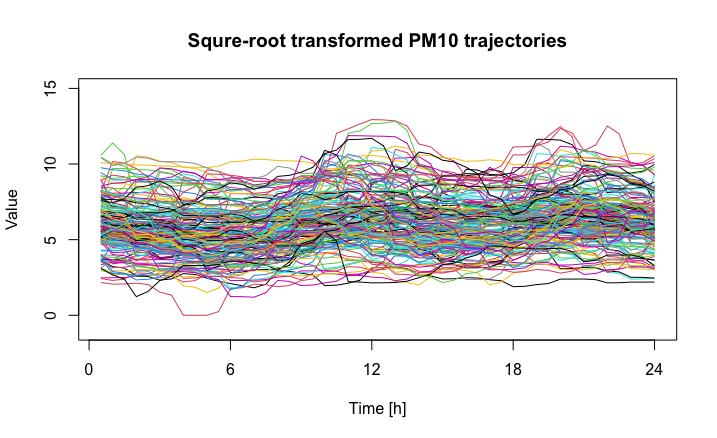}
	\caption{Square-root transformed PM10 concentration trajectories}
	\end{figure}	
Then the discrete vectors were transformed into functional objects using 10 cubic B-spline basis functions and the least squares fitting.  To remove weekly seasonality,  the mean functions were computed for each day of  the week and were used to centralize the trajectories before prediction. 

\subsubsection{Prediction of the smoothed PM10 concentration}
Suppose that the current time in a day is $\tau$, where $\tau\in(0,24]$, and there is partially-observed curve whose latter trajectory is to be predicted (i.e., the observation is only available over $[0,\tau]$ and the remainder of the curve on $(\tau,24]$ needs to be predicted). The one-step ahead prediction was conducted and the corresponding fFPE was computed. The minimum values of fFPE (over $d_x$ and $d_y$) are shown in Table 3 corresponding to different values of $p$ and $d$. Figure 3 shows the partial prediction of two randomly selected curves as $\tau=8\colon00,\ 12\colon00,\ 16\colon00$ respectively. Note that the prediction residual functions are not necessarily centered at zero, and thus the mean function needs to be adjusted when computing the intra-curve prediction. The final prediction is 
\[
\widehat{Y}_{n+1}|_{(\tau,1]}=\hat{\mu}_{n+1}|_{(\tau,1]}+\hat{\mu}_{e}|_{(\tau,1]}+\sum_{h=1}^p(\hat{\Phi}_h(Y_{n+1-h}-\hat{\mu}))|_{(\tau,1]}+\hat{\beta}_\tau(Y_{n+1}|_{[0,\tau]}-\widehat{Y}_{n+1}|_{[0,\tau]}-\hat{\mu}_{e}|_{[0,\tau]}),
\]
where $\hat{\mu}_{n+1}$ is the estimated weekday mean function, and $\hat{\mu}_{e}$ is the estimated mean function of the prediction residuals in the first step.
\begin{table}[h!]
	\centering
	\begin{small}
	\label{my-label}
	\begin{tabular}{|c|cccccccc|}
	\hline
	\hline
	\multicolumn{9}{|c|}{$\tau=8\colon00$}\\
	\hline
	&$d=1$ & $d=2$ & $d=3$ & $d=4$ & $d=5$&$d=6$ & $d=7$ & $d=8$ \\
	$p=0$&{\bf0.5993}&{\bf0.5993}&{\bf0.5993}&{\bf0.5993}&{\bf0.5993}&{\bf0.5993}&{\bf0.5993}&{\bf0.5993}\\
	$p=1$&0.6278& 0.6380& 0.6330& 0.6494& 0.6459& 0.6452& 0.6462& 0.6635\\
	$p=2$&0.6349& 0.6591& 0.6568& 0.6695& 0.6742& 0.6965& 0.7659& 0.7933\\
	$p=3$&0.6357& 0.6739& 0.6412& 0.6520& 0.6542& 0.7130& 0.7966& 0.8346\\
	\hline
	\multicolumn{9}{|c|}{$\tau=12\colon00$} \\
	\hline
	&$d=1$ & $d=2$ & $d=3$ & $d=4$ & $d=5$&$d=6$ & $d=7$ & $d=8$\\
	$p=0$&{\bf0.4184}&{\bf0.4184}& {\bf0.4184}& {\bf0.4184}& {\bf0.4184}& {\bf0.4184}& {\bf0.4184}&{\bf0.4184}\\
	$p=1$&0.4274& 0.4344& 0.4260& 0.4498& 0.4655& 0.4605& 0.4485& 0.4417\\
	$p=2$&0.4292& 0.4460& 0.4515& 0.4691& 0.4933& 0.4962& 0.5263& 0.5441\\
	$p=3$&0.4268& 0.4610& 0.4385& 0.4636& 0.4830& 0.5045& 0.5657& 0.5774\\	
	\hline
	\multicolumn{9}{|c|}{$\tau=16\colon00$} \\
	\hline
	&$d=1$ & $d=2$ & $d=3$ & $d=4$ & $d=5$&$d=6$ & $d=7$ & $d=8$\\
	$p=0$&0.1446& 0.1446& 0.1446& 0.1446& 0.1446& 0.1446& 0.1446& 0.1446\\
	$p=1$&0.1517& 0.1494& {\bf 0.1431}& 0.1472& 0.1444& 0.1447& 0.1525& 0.1514\\
	$p=2$&0.1519& 0.1490& 0.1436& 0.1494& 0.1453& 0.1474& 0.1717& 0.1744\\
	$p=3$&0.1514& 0.1510& 0.1535& 0.1625& 0.1580& 0.1675& 0.2004& 0.1925\\
         \hline 
         \hline
	\end{tabular}
	\caption{The minimum values of fFPE for different pairs of $p,d$, when $\tau=8\colon00$, $d_x=6$, $d_y=9$, $p=0$; when $\tau=12\colon00$, $d_x=7$, $d_y=8$, $p=0$; when $\tau=16\colon00$, $d_x=8$, $d_y=8$, $p=1$, $d=3$.}
	\end{small}
	\label{tab5}
\end{table}
	
\begin{figure}[!h]
	\centering
	\includegraphics[scale=0.4]{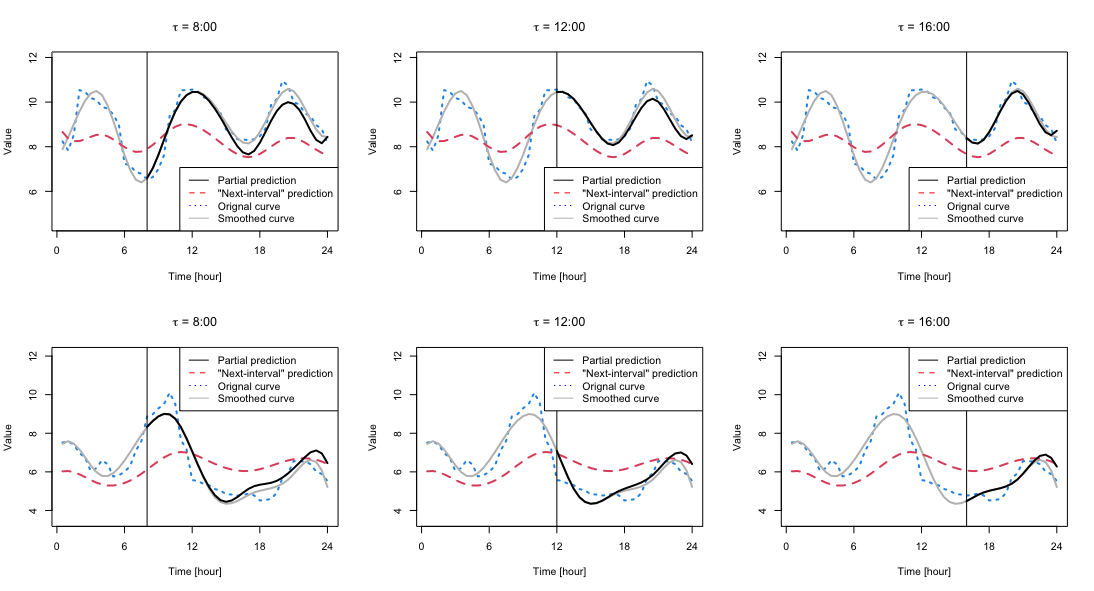}
	\caption{Partial functional prediction as $\tau=8:00,\ 12:00,\mbox{and}\ 16:00$}
\end{figure}

\subsubsection{Comparison with moving block method}
Shang (2017) proposed a functional time series prediction method, called the moving block method, to update the prediction with switching $\tau$.  The time support is shifted forward by $\tau$. Specifically, the $(\tau,1]$ block of the $m$-th curve is  combined with the $[0,\tau]$ block of the $(m + 1)$-th curve to form a new function. The new functions are a recombination of the original functional time series with the loss of the $[0,\tau]$ part of the first curve, which typically lays trivial effect on the prediction. The full-curve prediction method is then applied to the new functional time series, and the $[0,\tau]$ block of the predicted function is the updated prediction.
	
Table 4 includes the prediction MSE of the last 20 trajectories by PFP and the moving block method. It is noted that PFP robustly outperforms the moving block method over a broad range of values of $\tau$. The result is not unexpected since the moving block method actually belongs to ``next-interval'' prediction method, which provides complete curve prediction, while PFP aims to produce prediction only for the unobserved block, so the prediction error of the unobserved block provided by PFP should be smaller than that of the moving block method. Indeed one of the advantages of PFP over the moving block is that it directly uses the intra-curve variation, that is, the partially observed trajectory is directly treated as a part of the trajectory of interest rather than treating it artificially as part of the previous curve. A severe limitation of the moving block method is that the partially observed curve for the current trajectory is artificially forced to be a part of the previous one. This principle might work for some data settings but will not be reasonable for many other biological settings where the start of the curve has a well defined meaning (such as the onset of a stimulus presentation or a shock in biological experiments). 
	
\begin{table}[h!]
	\centering
	\begin{small}
	\label{my-label}
	\begin{tabular}{|p{1in}|p{0.6in}|p{0.6in}|p{0.6in}|}
	\hline
	\hline
	\multicolumn{1}{|c|}{Method} & \multicolumn{1}{c|}{$\tau=8\colon00$}& \multicolumn{1}{c|}{$\tau=12\colon00$}& \multicolumn{1}{c|}{$\tau=16\colon00$}\\
         \hline
         moving block& 0.56194 &0.34591 & 0.20138\\
         PFP method & {\bf 0.34789} & {\bf 0.26852} &  {\bf 0.10722}\\
         \hline
         \hline
	\end{tabular}
	\caption{Prediction MSE of the two methods.}
	\end{small}
\end{table}
	
\subsubsection{Prediction of the original trajectories}
Since the PM10 trajectories are not smooth and present seasonal dynamics, it is natural to implement the PFP method for the noisy case. The prediction results are compared with ARIMA model prediction, and PFP for smooth case was also implemented for comparison. We also applied linear interpolation when smoothing the original trajectories to incorporate the random errors, and then applied PFP for smooth case to finalize the prediction. 
	
The current time $\tau$ was assumed to be $12\colon00$, say the first 24 values were observed. The prediction methods were applied to predict the $h$-step ahead point values for the last 25 trajectories, where $1\le h\le 10$. Table 5 shows the prediction error of the three methods. Figure 4 shows part of the original centered time series and the corresponding smoothing residuals, and it is noted that after removing the smoothed functions, the residuals have no obvious seasonal trend compared with the original time series.

\begin{figure}[!h]
	\centering
	\includegraphics[scale=0.45]{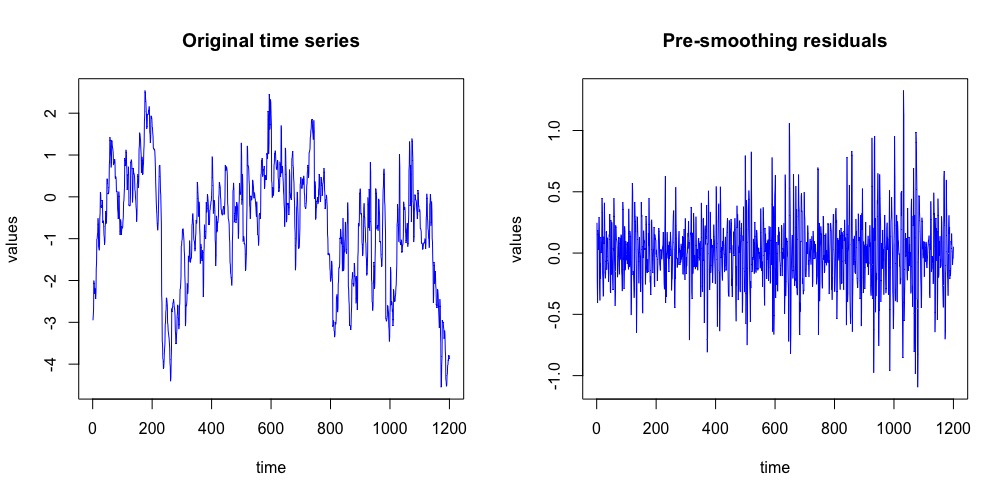}
	\caption{Part of the centered time series and the corresponding smoothing residuals.}
	\end{figure}
	\begin{table}[h!]
	\centering
	\begin{small}
	\label{my-label}
	\begin{tabular}{|c|c|c|c|c|}
	\hline
         \hline
         $h$ & $\mbox{PMSE}_n$ & $\mbox{PMSE}_s$ &$\mbox{PMSE}_a$& $\mbox{PMSE}_i$\\
         \hline
      	1& 0.1508980& 0.3540901& 0.2307175& 0.4642256\\
	2 & 0.2680703& 0.4757538& 0.6123183&0.6500848\\
	3 & 0.2309391& 0.2576980& 0.5273938&0.6132668\\
	4 & 0.4849306& 0.4972889& 0.8758383& 0.9419479\\
	5 & 0.3512944& 0.4108830& 0.9005394&0.9464275\\
	6 & 0.2363455& 0.3411949& 0.9953703&0.9732108\\
	7 & 0.2317724& 0.2619626& 0.9681887&0.9455279\\
	8 & 0.2184283& 0.2406333& 0.9389497&0.8657568\\
	9 & 0.2376993& 0.2154614& 0.9931003&0.8339264\\
	10& 0.1853883& 0.2210090& 1.1429001&0.9116115\\
         \hline
         \hline
	\end{tabular}
	\caption{Prediction MSE of the three methods (See Table 2 for the explanation of notations)} 
	\end{small}
\end{table}
	
Table 5 indicates that there exists dependence across the smoothing residuals since Step 4 and 5 significantly improve the prediction. These results demonstrate that PFP captures both the short-term dynamics (across smoothing residuals) and long-term dynamics (across and within smoothed functions). The ARIMA model can only give good predictions for the short-term predicted values but cannot give accurate predictions if we are interested in the long-term future. Linear interpolation does not perform well since the random errors contaminate the smooth signals and lead to bias in the estimated functional principal components.
	
\subsection{Analysis of traffic flow trajectories}
We now analyze the traffic flow data that was collected by a dual loop vehicle detector near the Shea-San Tunnel on National Highway 5 in Taiwan in 2009 (shared by Chiou (2012)). It refers to the vehicle count per minute over 15-min time intervals (96 observations for each day). There are 92 days of observed trajectories in total, and the goal is to predict the unobserved block of the last 12 trajectories. Figure 5 shows the raw daily trajectories and smoothed daily trajectories (smoothed with 21 B-splines).
\begin{figure}[!h]
\centering
\includegraphics[scale=0.4]{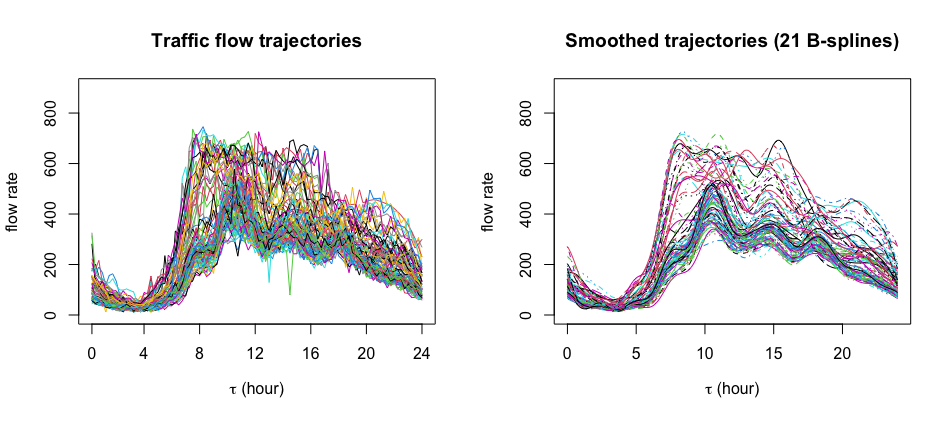}
\caption{Daily traffic flow trajectories and smoothed trajectories}
\end{figure}
Chiou (2012) proposed a functional mixture prediction method for independent trajectories which were first classified into several clusters. Intra-curve regression model within each cluster was applied to predict the unknown block in each potential cluster. The predictions in each cluster were combined to form the final prediction. Here, we used the first 80 trajectories as the training set to determine the cluster membership by subspace projection cluster 
algorithm (see Chiou and Li (2007)), and the last 12 trajectories are classified based on only the $[0,\tau]$ block.
	
In the testing set (the last 12 trajectories), for a sample $Y_k(t)$ observed up to $\tau$, we used the mean integrated prediction error 
(abbreviated as MIPE) to measure the performance of different methods
\[
\mbox{MIPE}(\tau)=\frac{1}{12}\sum_{k=1}^{12}\sqrt{\frac{1}{1-\tau}\int_{\tau}^1\{Y_{k+80}(t)|_{(\tau,1]}-\widehat{Y}^u_{k+80}(t)|_{(\tau,1]}\}^2{dt}}.
\]
Figure 6 shows the MIPE of the two methods. The result shows that proposed PFP method yields lower MIPE compared to the functional mixture prediction method. Although the functional mixture prediction method can work well in some cases, it has some limitations. First, the method classifies the future curve only based on the observed part, however, when the observed part is not very representative of the whole curve, the curve to be predicted is likely to be classified into a wrong cluster, which will potentially increase the prediction error. Furthermore, applying functional linear regression in each cluster actually reduces the sample size, resulting in larger estimation error. 
\begin{figure}[!h]
\centering
\includegraphics[scale=0.45]{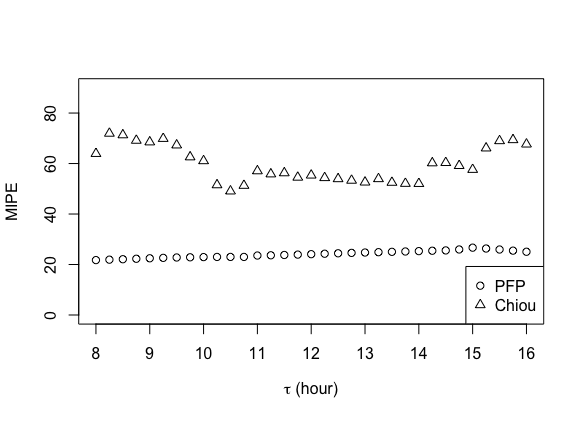}
\caption{MIPEs corresponding to different $\tau$ ranging from 8:00 to 16:00.}
\end{figure}

\section{Conclusion}
The contribution of this paper is a new functional prediction methodology that provides an update on the prediction given that the curve to be predicted is partially observed. The PFP method is motivated by the idea that the updated prediction should be a projection onto the $\sigma$-algebra expanded by the past observed trajectories and the partial observation. The prediction algorithm is a step-wise procedure, and can be applied to both smooth and non-smooth functions. In non-smooth case, the functional techniques can be applied for removing the seasonal trend, and univariate time series models can be applied to predict the smoothing residuals more effectively.

There are already several prediction methods for functions which we summarize here. In the functional time series prediction method (e.g.\ Aue et al.\ (2015)), 
the ``next-interval'' prediction only considers the big picture of the next function. In the setting where there is available partial observation, it is more natural to use all the available information (in particular, intra-curve information) to predict the unobserved part in order to  improve prediction. The primary limitation of the existing full-curve prediction methods is that they do not incorporate this available information. Another method, the moving block method (Shang, 2017), is essentially ``next-interval'' prediction method, so it has the same limitations discussed above. Another limitation of this method is that it is unnatural to arbitrarily assign starting and ending points of a curve especially in studies where such points are explicitly determined (e.g., start of the day; start of a trial in an experiment). As for the fully functional regression method (see e.g.\ Ramsay and Silverman (2005)), while it is commended for incorporating intra-curve information, its limitation is that it does not take into account the correlation across trajectories. This is a serious issue when the dependence across functions is non-trivial and the past trajectories are highly informative for predicting future trajectories. The method of Chiou et al.\ (2012) is an extension of functional regression which smartly and intuitively combines functional regression and clustering. The limitation of this approach is that when the partial observation does not give a strong indication of cluster membership, the classification will not be reliable and this could lead to serious prediction errors. It can be challenging to classify time series with low signal-to-noise ratio or with short time series length. Moreover, the sample size is potentially reduced (per cluster) as we need to do estimation for each cluster separately.  On the general approach of functional time series prediction after smoothing the curve by linear interpolation, although this is a way to jointly incorporate the information of both long-term and short-term dynamics, random errors will be included in the obtained curve and that can result in bias. 

The PFP method has several advantages. Since functional data are usually obtained in consecutive time intervals, the time series structure (e.g., autocorrelation) ubiquitously exists in functional data, and thus the PFP method has a broad range of applicability.  Additionally, it is flexible, in the sense that, users can decide whether or not the time series structure should be taken into account when predicting data. The PFP method is developed for predicting the unobserved block, so comparing with the full-curve prediction (including the moving block method), the prediction error of the unobserved block should be smaller. The simulation studies and real data analysis demonstrate that PFP consistently yields superior and competitive prediction.

\end{document}